\documentclass[showpacs,notitlepage,prd,aps,longbibliography,onecolumn]{revtex4-1}
\usepackage{graphicx}\usepackage{amsmath}\usepackage{amssymb}\usepackage{slashed}
\usepackage{color}
\usepackage{graphicx}
\newcommand{\Ref}[1]{(\ref{#1})}

\usepackage{environ}
\NewEnviron{myequation}{%
\begin{equation}
\scalebox{1.1}{$\BODY$}
\end{equation}}

\usepackage{booktabs} 
\usepackage{multirow}
\usepackage{soul} 
\usepackage{microtype}

\begin{document}
\title{Dispersion Forces Between Fields Confined to Half Spaces}

\author{M. Bordag$^a$ and
 I.G. Pirozhenko$^{b,c}$\\
$^a$\footnotesize{{\sl Institut f\"ur Theoretische Physik, Universit\"at Leipzig, Germany.}}\\
$^b$\footnotesize{{\sl Bogoliubov Laboratory of Theoretical Physics, JINR, Dubna, Russia }}\\
$^c$\footnotesize{\sl  Dubna State University,  Dubna, Russia}
}

\date{\small \today}
\begin{abstract}
We consider the Casimir effect for a scalar field interacting with another scalar field that is confined to two half spaces. This model is aimed to mimic the interaction of the photon field with matter in two slabs. We~use Dirichlet boundary conditions on the interfaces for the fields in the half spaces and calculate their one-loop contribution to the wave equation for the other field. We~perform the ultraviolet renormalization and develop a convenient formalism for the calculation of the vacuum energy in this configuration.
\end{abstract}

\maketitle







\section{Introduction}
In its initial formulation,  the Casimir effect is the attraction  between  conducting plates due to the vacuum fluctuations  of the electromagnetic field \cite{casi48-73-360}.
It was later generalized to the dispersion force between dielectric half spaces \cite{lifs56-2-73}. The evolution went on to include temperature of  the plates and dissipation (see, e.g.,  \cite{BKMM}). The material of the plates was modeled by dipoles and/or  polarization fields. An equivalent, perhaps more general, approach is macroscopic quantum electrodynamics~\cite{sche08-58,phil10-12-123008,hors14-16-013030}. It is a~common feature of these approaches that the resulting Hamiltonian is quadratic in the fields. First~papers going beyond this were   on the Casimir effect with respect to graphene~\cite{bord09-80-245406,pyat09-21-025506,fial12-27-1260007}. Here, the coupling is the usual electrodynamic one,  $\bar{\psi} \gamma^{\mu}\psi A_{\mu}$, which is a vertex with three lines, and~the corresponding Hamiltonian is no longer quadratic. However, since the spinor field describing the electrons in graphene is confined to a two-dimensional surface in   three-dimensional space, the reflection coefficients could be expressed explicitly in terms of the  polarization tensor of the electrons calculated within otherwise unconstrained (2 +1)-dimensional  field theory of the spinor field representing the electrons.  Now, if we assume two half spaces wherein a spinor or another field is confined, we have a (3 +1)-dimensional field theory restricted to two half spaces. We think that there is an upcoming interest in such type of calculations, which comes from the  high precision measurement of dispersion forces (up to femto Newton) and that there is a challenge   to account for internal dynamical properties of the interacting bodies in more detail. This means that,  in future, one will be tempted to calculate Casimir forces between fields inside the bodies, electron-hole, and phonon fields, for instance.

In the present paper,  we consider  the simplest  model of the mentioned type consisting of a  scalar field $\phi(x)$ in the whole space, mimicking the photon, and  a field $\psi(x)$ confined  to the half spaces $z<0$ and $z>L$, where $L$ is  the width of the gap between them, mimicking   matter. For the interaction of the fields, we take for simplicity
\begin{equation}
\label{e1}
  L_{int}(x)=\lambda \phi (x)\psi^2(x), \quad x_{\mu}=(x_{\alpha},z), \; \alpha=0,1,2
\end{equation}
where $\lambda$ is a coupling constant with a dimension of inverse length. We assume   boundary conditions on $\psi(x)$ at $z=0$ and at $z=L$ in order to have a mathematically consistent model. Later, in~applications, one would choose  appropriate boundary conditions. The interaction part of the action~is
\begin{equation}\label{e2}
  S_{int}(x)=\lambda \int d^3 x_{\alpha} \Bigl( \int\limits_{-\infty}^0 dz\, \phi (x)\psi^2(x)+
	\int\limits_{L}^{\infty} dz\, \phi (x)\psi^2(x) \Bigr), \quad \alpha=0,1,2.
\end{equation}

\textls[-10]{Our notations are taken from the relativistic  quantum field theory, so $x=(x^0, x_{||},z)$ is the 4-coordinate, and $x_{||}=(x,y)$ is the coordinate parallel to the surfaces. The same notations will also be   taken  for the momentum~variables.}  

The problem to be worked out is to get formulas that allow for effective numerical computation of the Casimir force {\color{red}or} the free energy. 
An obstacle to surmount is the UV-divergence appearing in the loop as well as the broken translational invariance in  the $z$-direction.
\section{Scalar Field Confined to Half Spaces and the Casimir Effect}
\vspace{-6pt}

\subsection{The Model}
Given a scalar field $\psi(x)$ confined to the half spaces $z<0$ and $z>L$, which fulfill 
 Dirichlet boundary conditions on $z=0$ and $z=L$. Assume interaction with another scalar field $\phi(x)$, defined in the whole space. In one-loop approximation,  the propagator of the field $\phi(x)$ is given by $\Delta^{-1}=\Delta_0^{-1}+\Pi$, where $\Delta_0$ is the wave operator and $\Pi$ is the polarization operator induced by the interaction with the field $\psi$ in half spaces.

In the lowest order perturbation theory, the equation of motion for the field $\phi(x)$ is
\begin{equation}
\label{a1}
\int d^4 x' \Bigl(-\partial^2_x \delta(x-x')+\Pi(x,x')\Bigr)\phi(x')=0
\end{equation}
with the polarization operator $\Pi(x,x')$
\begin{eqnarray}
\label{a1a}
\Pi(x{_\alpha};z,z')=-i\lambda^2 D_{\mathcal{D}}(x_{\alpha};z,z')^2=-i\lambda^2 \; \parbox[c]{30mm}{\includegraphics[width=3cm]{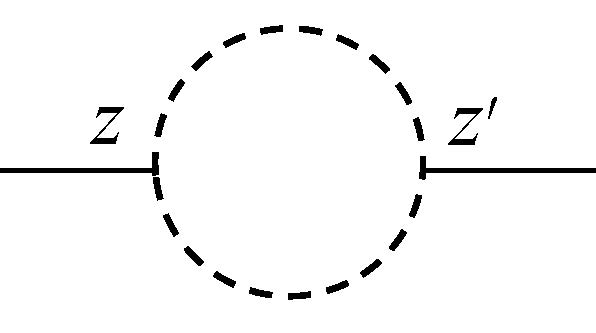}} \, 
\end{eqnarray}
given by the one loop scalar field diagram. The solid line corresponds  to the field $\phi$ and the dashed line to  the field $\psi$.

Using the translation invariance in all directions except the $z$-direction, we rewrite~Equation \Ref{a1} as
\begin{equation}
\label{a2}
\int d z' \Bigl((\Gamma^2+\partial^2_z) \delta(z-z')+\Pi_{\Gamma}(z,z')\Bigr)\phi(z')=0
\end{equation}
with
\begin{equation}
\label{a3}
\phi(z)=\int d^3 x_{\alpha} e^{i k_{\alpha}x_{\alpha}} \phi(x), \quad \alpha=0,1,2
\end{equation}
and
\begin{equation}
\label{a4}
\Pi_\Gamma(z,z')=\int  {d^3 x_{\alpha}}   e^{i k_{\alpha}x_{\alpha}} \, \Pi(x_{\alpha};z,z')
\end{equation}
where we used
\begin{equation}
\label{a5}
\Pi(x,x')\equiv \Pi(x_{\alpha}-x_{\alpha}^{\prime};z,z')
\end{equation}
and $\Gamma=\sqrt{k_{\alpha}k^{\alpha}+i 0}$.

In Equation \Ref{a1a}, $D_{\mathcal D}$  is the propagator of the field $\psi$ obeying Dirichlet boundary conditions.
The Fourier transform in the translational invariant directions is as follows:
 \begin{equation}
\label{a7}
\Pi_{\Gamma}(z,z')=\lambda^2\int d^3 x_{\alpha} e^{i k_{\alpha}x_{\alpha}} D_{\mathcal D}(x_{\alpha};z,z')^2
\end{equation}
where the same notation as in Equation \Ref{a5} is used, which is
\begin{equation}
\label{a8}
D_{\mathcal D}(x,x')\equiv D_{\mathcal D}(x_{\alpha}-x_{\alpha}^{\prime};z,z').
\end{equation}
Then, a single line of the field $\psi$ is
\begin{equation}
\label{a9}
D_{\mathcal D}(x,x')=\sum_{\sigma=\pm1}\sigma D(x_{\alpha}-x_{\alpha}^{\prime};z-\sigma z'),
\end{equation}
and
\begin{equation}
\label{a10}
D(x)= D(x_{\alpha};z)= \int\frac{d^4 q}{(2\pi)^4}\frac{e^{-i q x}}{q^2-m^2+i 0}
\end{equation}
is the usual propagator of the scalar field $\psi$.
%
\subsection{Transition to $TGTG$ Formula}
Understanding $\Pi(x,x')$ in Equation \Ref{a1} as a potential, $V(x,x')$, we  use the so called $TGTG$  
 formula where $T$ and $G$ stand  for $T$-matrix and Green's functon, accordingly.
 Employing Equations (3.112) and (10.40) in \cite{BKMM}, we obtain   the vacuum energy of the field $\phi $  in the presence of the half spaces: 
\begin{equation}
\label{a11}
E=-\frac{i}{2}\int\frac{d^3 k_{\alpha}}{(2 \pi)^2}{\mbox Tr} \ln(1-{\mathcal M})
\end{equation}
where
\begin{equation}
\label{a12}
{\mathcal M}(y,y')=\Pi_1(y,y_1)G_0(y_1-z_1)\Pi_2(z_1,z_2)G_0(z_2-y')
\end{equation}
with
\begin{equation}
\label{a13}
G_0(z)=\int\frac{d k_3}{2\pi}\frac{e^{i k_3 z}}{\Gamma^2-k_3^2+i 0}=\frac{e^{i\Gamma|z|}}{-2 i \Gamma}
\end{equation}
being the free space Green's function, i.e., the propagator of the field $\phi$.  
In Equation \Ref{a11}, it is assumed that the Fourier transform in the $\alpha$-directions $(\alpha=0,1,2)$ is taken, and we use Equation \Ref{a4} for $\Pi_i$,
and
\begin{equation}
\label{a15}
G_0(z)=\int\frac{d^3 x_{\alpha}}{(2\pi)^3}e^{i k_{\alpha}x_{\alpha}}G_0(x_{\alpha};z),
\end{equation}
which results in Equation \Ref{a13}.

In Equation \Ref{a12}, we took as convention arguments $y,y_1,y'$ in the left half space and $z_1,z_2 $ in the right half space.
Integrations $\int_{-\infty}^0dy$ and $\int_L^\infty dz$ are assumed.
The $\Pi_1$ and $\Pi_2$ correspond to the potentials $T^A$ and $T^B$ in Equation~(10.40) in \cite{BKMM}. Specifically, $\Pi_1(y,y')$ is from the field in $z<0$ (left half space). Therefore, we set $\Pi_1(y,y')=0$ for $y>0$ or $y'>0$. The other one, $\Pi_2(z,z')$, is from the field $\psi$ in the right half space, with  $z\le L$ and $z'\le L$, and we set  $\Pi_2(z,z')=0$ for $z<L$ or $z'<L$.
Further, we use the notation $\Pi_{\Gamma}(z,z')$, as given by Equation \Ref{a7}, which goes without index to represent   $\Pi_1$ and $\Pi_2$, entering Equation \Ref{a12}:
\begin{equation}
\label{a18}
\Pi_1(y,y')=\Pi_{\Gamma}(y,y')=\Pi_{\Gamma}(-y,-y'), \quad \Pi_2(z,z')=\Pi_{\Gamma}(z-L,z'-L).
\end{equation}

The polarization tensor $\Pi_2(z,z')$ is defined for the interface being located in $z=L$. Thus, we need in Equation \Ref{a7} the propagator $D_D$ with Dirichlet boundary conditions at $z=L$. Therefore, in place of Equations \Ref{a8}  and \Ref{a9}, we have
\begin{equation}
\label{a17}
D_{{\mathcal D}_L}(x_{\alpha}-x^{\prime}_{\alpha};z,z')=\sum_{\sigma=\pm1}\sigma D(x_{\alpha}
-x_{\alpha}^{\prime};z-L-\sigma (z'-L))=D_{{\mathcal D}}(x_{\alpha}-x^{\prime}_{\alpha};z-L,z'-L).
\end{equation}

$D_{D_L}$ denotes the propagator with boundary conditions on $z=L$, and $D_{\mathcal D}$ denotes the propagator with boundary conditions at $z=0$. The relation between them is given by
Equation~\Ref{a17}.

\textls[-5]{With these remarks, $\mathcal{M}$,  Equation \Ref{a18}, can be rewritten as (accounting for the trace in Equation~\Ref{a11})}
\begin{equation}
\label{a19}
{\mathcal M}={\mathcal N}_1 \cdot {\mathcal N}_2
\end{equation}
\begin{eqnarray}
\label{a20}
{\mathcal N}_1=\int\limits_{-\infty}^{0}d y  \int\limits_{-\infty}^{0}d y'  \, \frac{e^{-i \Gamma(y+y')}}{-2 i \Gamma}\,\Pi_1(y,y')
=\int\limits_0^{\infty}d y \int\limits_0^{\infty}d y' \, \frac{e^{i \Gamma(y+y')}}{-2 i \Gamma} \,\Pi_{\Gamma}(y,y')\\
\label{a21}
{\mathcal N}_2
=\int\limits_
L^{\infty}d z \int\limits_L^{\infty}d z' \,  \frac{e^{i \Gamma(z+z')}}{-2 i \Gamma} \, \Pi_2(z,z')=
\int\limits_
L^{\infty}d z \int\limits_L^{\infty}d z' \, \frac{e^{i \Gamma(z+z')}}{-2 i \Gamma}\,\Pi_{\Gamma}(z-L,z'-L).
\end{eqnarray}

Doing the substitutions, $z\to z +L$, $z'\to z'+L$ thus gives
\begin{equation}
\label{a22}
{\mathcal N}_2=e^{2 i \Gamma L}\int\limits_0^{\infty}d z \int\limits_0^{\infty}d z' \, \frac{e^{i \Gamma(z+z')}}{-2 i \Gamma}\,\Pi_{\Gamma}(z,z').
\end{equation}

This way, we can define
\begin{equation}
\label{a23}
{\mathcal N}=\int\limits_0^{\infty}d z \int\limits_0^{\infty}d z' \, \frac{e^{i \Gamma(z+z')}}{-2 i \Gamma}\,\Pi_{\Gamma}(z,z') 
\end{equation}
and ${\mathcal N}_1={\mathcal N}$ and ${\mathcal N}_2=e^{2 i \Gamma L}{\mathcal N}$ hold. As a result, Equation \Ref{a11} turns into
\begin{equation}
\label{a24}
E=-\frac{i}{2}\int\frac{d^3 k_{\alpha}}{(2 \pi)^2}{\mbox Tr} \ln(1-{\mathcal N}^2 e^{2 i \Gamma L}),
\end{equation}
and ${\mathcal M}={\mathcal N}^2 e^{2 i \Gamma L}$.
Comparing Equation \Ref{a24} with the Lifshitz formula at zero temperature, one can define the reflection coefficient of the half spaces in terms of the factors $\mathcal N$:
\begin{equation}
r(\omega,k_{||};\lambda, m\bigr)=\mathcal N\bigl(\sqrt{\omega^2-k_{||}^2}; \lambda, m\bigr) 
\end{equation}
where we indicated the variables and parameters that $\cal N$ depend  on.

\section{Polarization Operator in Half Space}
In the preceding section, we have defined Equation \Ref{a23}  for the factors ${\mathcal N}$, entering the vacuum energy (Equation \Ref{a24}), and introduced $\Pi$ by Equation \Ref{a1a}.
\textls[-15]{The propagator of the field $\psi$ obeys  Dirichlet boundary conditions at $z=0$ and is given by Equations \Ref{a9} and \Ref{a10}.
Inserting it into Equation \Ref{a1a}, we~obtain}
\begin{eqnarray}
\label{a28}
\Pi(z_{\alpha};z,z')=-i\lambda^2\int \frac{d^4 q}{(2\pi)^4} \int \frac{d^4 q'}{(2\pi)^4} \sum_{\sigma,\sigma'}\sigma \sigma' \frac{e^{-i(q_{\alpha}+q'_{\alpha})z_{\alpha}+i q_3(z-\sigma z')+i q'_3(z-\sigma' z')}}{(-q^2+m^2-i 0)(-q^{\prime 2}+m^2-i 0)}
\end{eqnarray}
where $\sigma$ and $\sigma'$ take values $\pm1$. With Equation \Ref{a7} and the integral representation of the $\delta$-function,
$$\int d^3 z_{\alpha}e^{i (k_{\alpha}-q_{\alpha}-q'_{\alpha})z_{\alpha}}=(2\pi)^3 \delta^{3}(k_{\alpha}-q_{\alpha}-q_{\alpha}^{\prime}),$$ we can integrate with respect to $q'_{\alpha}$ and arrive at
\begin{myequation}
\label{a29}
\Pi_{\Gamma}(z,z')
=-i\lambda^2  \sum\limits_{\sigma,\sigma'} \sigma \sigma' \int \frac{d^3 q_{\alpha}}{(2\pi)^3} \int \frac{d q_{3}}{2\pi} \int \frac{d q_{3}^{\prime}}{2\pi} \frac{e^{i q_3(z-\sigma z')+i q'_3(z-\sigma' z')}}{(-q_{\alpha}^2+q_3^2+m^2 -i 0)(-(k_{\alpha}-q_{\alpha})^{2}+q_3^{\prime 2}+m^2-i 0 )} 
\end{myequation}
where $\Gamma=\sqrt{k_{\alpha} k^{\alpha}+i 0}$ .

To proceed, we divide $\Pi_{\Gamma}(z,z')$ into the translational invariant part, $\Pi_{\Gamma}^{(t)}(z,z')$, arising from $\sigma=\sigma'=+1$ in the sum of Equation \Ref{a29}, and the remaining part,   $\Pi_{\Gamma}^{(nt)}(z,z')$:
\begin{equation}
\label{a30}
\Pi_{\Gamma}(z,z')=\Pi_{\Gamma}^{(t)}(z,z')+\Pi_{\Gamma}^{(nt)}(z,z').
\end{equation}
This will be treated separately.

We start from the translational non-invariant part,  $\Pi_{\Gamma}^{(nt)}(z,z')$. Here, we perform the Wick rotation, $q_0\to i q_4$ and use the $\alpha$-representation (parametric representation) with parameters $\alpha_1$ and $\alpha_2$ for the factors in the denominator of Equation \Ref{a29}. After that, the momentum integration can be carried out {\color{red}} and we arrive at
\begin{equation}
\label{a31}
\Pi_{\Gamma}^{(nt)}(z,z')=\frac{\lambda^2}{(4\pi)^{5/2}}\mathop{{\sum}'}_{\sigma,\sigma'=\pm1}\sigma \sigma' \int \frac{d\alpha_1 d\alpha_2}{\sqrt{\alpha_1\alpha_2}} \frac{e^{-H-A}}{(\alpha_1+\alpha_2)^{3/2}}
\end{equation}
where the prime over the sum sign means that the term with $\sigma=\sigma'=1$ is excluded, and
\begin{eqnarray}
\label{a32}
&&A=\frac{(z-\sigma z')^2}{4\alpha_1}+\frac{(z-\sigma' z')^2}{4\alpha_2}=\frac{1}{4}\left\{\Bigl(\frac{1}{\alpha_1}+\frac{1}{\alpha_2}\Bigr)(z^2+z^{\prime 2})-\Bigl(\frac{\sigma}{\alpha_1}+\frac{\sigma'}{\alpha_2}\Bigr) 2 z z'\right\}\\ \label{a33a}
&&H=\frac{\alpha_1\alpha_2}{\alpha_1+\alpha_2}\gamma^2+(\alpha_1+\alpha_2)m^2, \quad \gamma=\sqrt{q_4^2+q_1^2+q_2^2}.
\end{eqnarray}

After the change of variables, $\alpha_1=s x$ and $\alpha_2=s(1-x)$, one integration can be performed here,  and we obtain
\begin{equation}
\label{a33}
\Pi_{\Gamma}^{(nt)}(z,z')=\frac{\lambda^2}{32\pi^2}\mathop{{\sum}'}_{\sigma,\sigma'=\pm1}\sigma \sigma' \int \limits_0^1 \frac{d x}{\sqrt{x(1-x)}} \frac{e^{-2 \sqrt{\tilde{A}\tilde{H}}}}{\sqrt{\tilde{A}}}
\end{equation}
where
\begin{equation}
\label{a34}
\tilde{A}=\frac{1}{4}\left\{\Bigl(\frac{1}{x(1-x)}\Bigr)(z^2+z^{\prime 2})-\Bigl(\frac{\sigma}{x}+\frac{\sigma'}{1-x}\Bigr) 2 z z'\right\}, \quad
\tilde{H}=x(1-x)\gamma^2+m^2.
\end{equation}

We insert this expression into Equation \Ref{a23} and obtain the translational non-invariant part of the factor  ${\mathcal N}$,
\begin{equation}
\label{a35}
{\mathcal N}^{(nt)}=\frac{\lambda^2}{64\pi^2\gamma}\int\limits_0^{\infty}d z \int\limits_0^{\infty}d z' \mathop{{\sum}'}_{\sigma,\sigma'=\pm1}\sigma \sigma'
\int \limits_0^1 \frac{d x}{\sqrt{x(1-x)}}\frac{1}{\sqrt{\tilde{A}}}
\, e^{- \gamma(z+z')-2 \sqrt{\tilde{A}\tilde{H}}}.
\end{equation}

In order to simplify the integration over $z$ and $z'$,   we first turn the $(z,z')$ - plane, $z_{\pm}=z\pm z'$, and then substitute $z_{-}=\mu z_{+}$. After that, the integration with respect to $z_{+}$ yields
\begin{equation}
\label{a36}
{\mathcal N}^{(nt)}=\frac{\lambda^2}{128\pi^2\gamma}\mathop{{\sum}'}_{\sigma,\sigma'=\pm1}\sigma \sigma' \int \limits_0^1 \frac{d x}{\sqrt{x(1-x)}}
\int\limits_{-1}^{1}
\frac{d\mu}{\sqrt{A_{\sigma,\sigma'}}}\frac{1}{\gamma+2\sqrt{\tilde{H} A_{\sigma,\sigma'}}}
\end{equation}
where $\tilde{H}$ is given in Equation \Ref{a34}, and
\begin{equation}
\label{a37}
A_{\sigma,\sigma'}=\frac{1}{8}\left\{\left[\frac{1-\sigma}{x}+\frac{1-\sigma'}{1-x}\right]+\left[\frac{1+\sigma}{x}+\frac{1+\sigma'}{1-x}\right]\mu^2\right\},\quad \sigma, \sigma'=\pm1.
\end{equation}

 Finally, we have
\begin{equation}
{\mathcal N}^{(nt)}=N_{--}+2 N_{-+}
\label{a38}
\end{equation}
with $N_{--}$ and $N_{-+}$ corresponding respectively  to the terms with $\sigma=\sigma'=-1$ and with $\sigma=1$ and $\sigma'=-1$ 
 in the sum of Equation \Ref{a36}: 
\begin{eqnarray}
\label{a39}
&&N_{--}=-\frac{\lambda^2}{32\pi^2\gamma m^2}\left[\frac{\gamma}{6}-\int\limits_0^1 d x \sqrt{x(1-x)} \sqrt{x(1-x)\gamma^2+m^2}\right] \\
\label{a40}
&&N_{-+}=-\frac{\lambda^2}{64\pi^2\gamma}\int\limits_{-1}^{1}d\mu\int\limits_0^1\frac{d x}{\sqrt{\mu^2 x+(1-x)}}
\frac{1}{\left(\gamma+\sqrt{\left(\gamma^2+\frac{m^2}{x(1-x)}\right)\left(\mu^2 x+(1-x)\right)}\right)}.
\end{eqnarray}

The remaining integrals can be easily evaluated numerically.
The asymptotics of ${\mathcal N}^{(nt)}$ are
\begin{eqnarray}\label{a41}
{\mathcal N}^{(nt)}|_{\gamma\to0}=\frac{\lambda^2}{128\pi^2\gamma}\left\{-\frac{\pi}{2 m} +\frac{ 4\gamma}{3 m^2}+{\mathcal O}(\gamma^2)\right\} \\
\label{a42}
{\mathcal N}^{(nt)}|_{\gamma\to\infty}=\frac{\lambda^2}{128\pi^2\gamma}\left\{-\frac{1.28987}{\gamma}+{\mathcal O}(1/\gamma^2)\right\}.
\end{eqnarray}

The translational invariant part $\Pi_{\Gamma}^{(t)}$, defined in Equation \Ref{a30},  can be obtained from    Equation \Ref{a1a} by  dropping  the index `${\mathcal{D}}$' in the propagator, which is equivalent to considering the  term with $\sigma=\sigma'=1$  in Equation \Ref{a28}. We denoted  this term by $\Pi_{++}$. It has full 4-dimensional symmetry and is a function of $x-x'$:
\begin{eqnarray}
\label{a43}
\Pi_{++}(x)=-i\lambda^2 D(x)^2.
\end{eqnarray}

 The 4-dimensional Fourier transform  of $\Pi_{++}(x) $ has the known parametric representation 
\begin{equation}
\label{a44}
\Pi_{++}(q^2)=\lambda^2 \int \frac{d \alpha_1 d \alpha_2}{(4\pi(\alpha_1+\alpha_2))^{\frac{3}{2}-\varepsilon}}\exp\left(-\frac{\alpha_1\alpha_2}{\alpha_1+\alpha_2}q^2-(\alpha_1+\alpha_2)m^2\right), \quad q^2=\gamma^2+q_0^2
\end{equation}
where the Wick rotation is performed. In Equation \Ref{a44}, we introduced $\varepsilon$ as the parameter of  the dimensional regularization.  Further, we perform the renormalization in a way where $\Pi_{++}^{ren}(q^2=0)=0$ holds. This ensures that the mass of the field $\psi$ does not change. Technically, we achieve this by
\begin{equation}
\label{a45}
\Pi_{++}^{ren}(q^2)=\left.\left(\Pi_{++}(q^2 \delta)-\Pi_{++}(q^2 \delta)|_{\delta=0}\right)\right|_{\delta=1}
\end{equation}
where $\delta$ is an auxiliary parameter.
Next we need $\Pi_{++}(z-z')$, defined in Equation \Ref{a4}. Using
\begin{equation}
\label{a46}
\Pi_{++}^{ren}(x)=\int\frac{d^4q}{(2\pi)^4}e^{-i q_{\mu}x^{\mu}}\Pi_{++}(q^2)
\end{equation}
we obtain
\begin{eqnarray}
\label{a47}
\Pi_{++}^{ren}(z-z')&=&\int d^3 x_{\alpha}e^{i k_{\alpha}x^{\alpha}}\Pi_{++}^{ren}(x)\\
&=&\int\frac{d q_3}{2\pi}e^{i q_3(z-z')}\Pi_{++}^{ren}(q^2). \nonumber
\end{eqnarray}

Inserting Equation \Ref{a47} into Equation \Ref{a23}, we arrive at
\begin{equation}
\label{a48}
{\mathcal N}^{(t)}_{ren}=\int\limits_0^{\infty}d z \int\limits_0^{\infty}d z' \, \frac{e^{-\gamma(z+z')}}{2 \gamma}\,\int\frac{d q_3}{2\pi}e^{i q_3(z-z')}\Pi_{++}^{ren}(q^2)
\end{equation}
where the integration over $z$ and $z'$ is easily carried out:
\begin{equation}
\label{a49}
\int\limits_0^{\infty}d z \int\limits_0^{\infty}d z' \, e^{-\gamma(z+z')} e^{i q_3(z-z')}=\frac{1}{\gamma^2+q_3^2}\equiv\int\limits_0^\infty d\alpha_3 e^{-\alpha_3(\gamma^2+q_3^2)}.
\end{equation}
The new, proper time representation with the parameter $\alpha_3$ is then used.  After integration over $q_3$, we finally obtain
\begin{eqnarray}
\label{a50}
{\mathcal N}^{(t)}=\frac{\lambda^2}{2^6 \pi^{5/2}\gamma}\int\frac{d \alpha_1 d \alpha_2 d\alpha_3}{(\alpha_1+\alpha_2)^{3/2-\varepsilon}}\frac{1}{\sqrt{D}}
\exp\left\{-\frac{D}{\alpha_1+\alpha_2}\gamma^2-(\alpha_1+\alpha_2)m^2\right\}
\end{eqnarray}
where $D=\alpha_1 \alpha_2 \delta+\alpha_3(\alpha_1+\alpha_2)$.
Further we'll omit the parameter $\varepsilon$ in the formulas, bearing in mind that  the subtraction of Equation \Ref{a45} is performed under the sign of the integration. In order to do the subtraction, we divide the integration area in Equation \Ref{a50} into sectors $\alpha_i<\alpha_j<\alpha_k$, following, e.g., \cite{Zavialov90}, p.134. Owing to the symmetry $\alpha_1\leftrightarrow \alpha_2$ in Equation \Ref{a50}, we have to account for three distinct sectors~\cite{Zavialov90}:
\begin{equation}
\label{a51}
 \begin{aligned}
\mbox{1 sector:}&& \alpha_1<\alpha_2<\alpha_3  \\
\mbox{2 sector:}&& \alpha_1<\alpha_3<\alpha_2  \\
\mbox{2 sector:}&& \alpha_3<\alpha_1<\alpha_2. 
 \end{aligned}
\end{equation}

Then, we change the variables,
\begin{equation}
t_i=\alpha_i/\alpha_{i+1} (i=1,2), \quad t_{n}=\alpha_n, \quad J=t_2 t_3^2
\end{equation}
and obtain
\begin{equation}
\label{a52}
\begin{array}{llllll}
\mbox{1 sector:}&\alpha_1=t_1 t_2 t_3, &\alpha_2=t_2 t_3,&\alpha_3=t_3,&D=t_3^2 t_2 d_1,&d_1=t_1 t_2 \delta+t_1+1 \\[0.2cm]
\mbox{2 sector:}& \alpha_1=t_1 t_2 t_3,&\alpha_2= t_3,&\alpha_3=t_2 t_3,&D=t_3^2 t_2 d_2,&d_2= t_1 t_2+t_1 \delta+1 \\[0.2cm]
\mbox{2 sector:}& \alpha_1=t_2 t_3,&\alpha_2=t_3,& \alpha_3=t_1 t_2 t_3,&D=t_3^2 t_2 d_2,&d_3= t_1 t_2+t_1+ \delta.
\end{array}
\end{equation}

Now the translational invariant part ${\mathcal N}^{(t)}$ comprises the contributions from six sectors, and half of them are equivalent. Thus, allowing for the multiplicity of the sectors,
one can  write
\begin{equation}\label{a53}
{\mathcal N}^{(t)}=2\sum_{i=1}^{3}{\mathcal N}^{(i)}
\end{equation}
where ${\mathcal N}^{(i)}$ corresponds to the contribution from  the  i-th sector defined in Equation \Ref{a51}. Integrating with respect to $t_3$,  we arrive at
\begin{equation}
\label{a54}
{\mathcal N}^{(1)}=\frac{\lambda^2}{2^6 \pi^{2}\gamma}\int\limits_{0}^1 \frac{d t_1}{t_1+1} \int\limits_{0}^1 \frac{d t_2}{t_2\sqrt{d_1}}\frac{1}{\sqrt{d_1 \gamma^2+t_2(t_1+1)^2m^2}}
\end{equation}
\begin{equation}
\label{a55}
{\mathcal N}^{(2)}=\frac{\lambda^2}{2^6 \pi^{2}\gamma}\int\limits_{0}^1 d t_1 \int\limits_{0}^1 \frac{d t_2 \sqrt{t_2}}{(1+t_1 t_2)\sqrt{d_2}}\frac{1}{\sqrt{t_2 d_2 \gamma^2+(t_1 t_2+1)^2m^2}}
\end{equation}
\begin{equation}
\label{a56}
{\mathcal N}^{(3)}=\frac{\lambda^2}{2^6 \pi^{2}\gamma}\int\limits_{0}^1 d t_1 \int\limits_{0}^1 \frac{d t_2 \sqrt{t_2}}{(1+t_1 )\sqrt{d_3}}\frac{1}{\sqrt{t_2 d_3 \gamma^2+(t_2+1)^2m^2}}.
\end{equation}

The UV divergence is sitting in the first sector at $t_2=0$. It is logarithmic as expected, and we have to do the subtraction of Equation \Ref{a45}, which does not cause any problems in our case.
The integrals can be evaluated numerically. The asymptotic expansion of  ${\mathcal N}^{(t)}$ reads
\begin{eqnarray}\label{a57}
{\mathcal N}^{(t)}|_{\gamma\to0}&=&\frac{\lambda^2}{64\pi^2\gamma}\left\{-\frac{0.7853}{m} +{\mathcal O}(\gamma^2)\right\}\\
\label{a58}
{\mathcal N}^{(t)}|_{\gamma\to\infty}&=&\frac{\lambda^2}{64\pi^2\gamma}\left\{-2\frac{\ln(\gamma/m)}{\gamma}
+\frac{0.6137}{\gamma}+{\mathcal O}(1/\gamma^2)\right\}.
\end{eqnarray}

Taking the parts of Equation \Ref{a23} corresponding to Equation \Ref{a30}, namely Equations \Ref{a41}, \Ref{a42}, \Ref{a57}, and \Ref{a58}, together, we obtain for the factor, entering Equation \Ref{a24},
\begin{eqnarray}\label{a58a}
{\mathcal N}|_{\gamma\to0}&=&\frac{\lambda^2}{128\pi^2\gamma}\left\{-\frac{\pi}{m} +\frac{4\gamma}{3m}+{\mathcal O}(\gamma^2)\right\}\\
\label{a59}
{\mathcal N}|_{\gamma\to\infty}&=&\frac{\lambda^2}{128\pi^2\gamma}\left\{-4\frac{\ln(\gamma/m)}{\gamma}
-\frac{0.0624567}{\gamma}+{\mathcal O}(1/\gamma^2)\right\}.
\end{eqnarray}
The factor $ {\mathcal N}$ as a function of momenta $\gamma$  is shown in Figure \ref{F1} . 

\section{Vacuum Energy}\label{sec4}
In the present model, the vacuum energy is defined by Equation \Ref{a24} and can be rewritten  in the form
\begin{equation}
\label{a60}
E=\frac{1}{4\pi}\int\limits_0^{\infty}d \gamma \gamma^2  \ln(1-{\mathcal N}^2 e^{-2 \gamma L}).
\end{equation}

For $\gamma\to0$, because of ${\mathcal N}^2\sim\frac{1}{\gamma^2}$ (see Equations \Ref{a41} and \Ref{a57}), the argument of the logarithm becomes negative and the vacuum energy acquires an imaginary part, which signifies some instability within the considered model.
It is known that an imaginary part of the effective action signals particle creation. Specifically,  in the Casimir--Polder interaction of polarizable dipoles, such instability signals the breakdown of the dipole approximation (see, for example, \cite{berm14-89-022127}, Equation (26) and subsequent discussion, and \cite{bord17-96-062504}, the section after Equation (142) concerning atom--wall interaction. As the integration in~Equation \Ref{a60} goes from zero to infinity, the formula yields a complex vacuum energy of the field $\phi$ for any finite width of the gap between the half spaces.


In fact, our model is aimed to mimic the 
interaction of the photon field with the electron and phonon fields in a solid.
 As is known, the Coulomb interaction between the electrons and the phonons is screened,
and the electron charge density interacts with the gradient of the phonon displacement field. This mechanism is described in many solid state textbooks (see, for instance,~\cite{Martin2004}). A characteristic feature of the model is the gradient in the interaction vertex, which turns into a momentum after Fourier transform. To account for this gradient in some way, we make the coupling momentum~dependent:
\begin{equation} \label{a61}
\lambda\to\lambda(\gamma)=\lambda_0 \sqrt{\gamma},
\end{equation}
 such that
 \begin{equation} \label{a62}
{\mathcal N}|_{\gamma\to0}\sim C, \quad {\mathcal N}|_{\gamma\to\infty}\sim \frac{\ln(\gamma)}{\gamma}.
\end{equation}

It is clear that this approximation is very crude, but we do not intend  to describe the real electron--phonon interaction, but rather the possibility to perform calculations with such a model. As~stated in the introduction, we consider the most simple model to develop the methods,  which may be helpful to  account for the contribution of the electron--phonon interaction to the Casimir force.

It should be mentioned that the factors ${\mathcal N}$ play the role of   reflection coefficients within the present approach for the $\phi$ field. Now, without the substitution according to Equation \Ref{a61}, these factor's asymptotics 
 are given by Equations \Ref{a54} and \Ref{a55}, and, with Equation \Ref{a61}, they behave as Equation \Ref{a62}.

For the model after the substitution of Equation \Ref{a61}, we consider the Casimir (vacuum) energy. Its~behavior for large separation can be obtained by scaling $\gamma\to\gamma/L$ in Equation \Ref{a60}. We obtain the factor $L^{-3}$ in front and ${\mathcal N}(\gamma/L)$ in the logarithm, which, for $L\to\infty$, is then given by Equation \Ref{a62}. After the substitution of Equation \Ref{a61}, we obtain
\begin{equation}
\label{a63}
E|_{L\to\infty}=\frac{1}{4\pi L^3}\int\limits_0^{\infty}d \gamma \gamma^2  \ln(1-\lambda^2 C_1 e^{-2 \gamma })=-\frac{1}{16 \pi L^3}\mbox{Li}_4(\lambda^2 C_1).
\end{equation}

Figure \ref{F2} shows ratio $\eta$ of the Casimir energy of Equation \Ref{a60} as well as the Casimir energy of the massless scalar field with Dirichlet boundary conditions on the plates, given by $E_{D}=- \pi^2/(1440 L^3)$ in the units $\hbar=c=1$.  At large separations, the ratio tends to a constant  determined by Equation \Ref{a63}.

\section{Conclusions}
\begin{figure}[h]
\centering
\includegraphics[width=7cm]{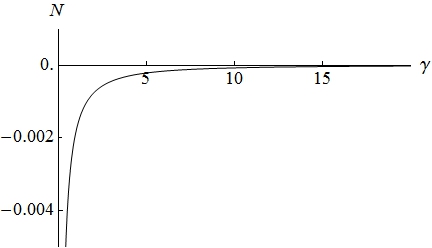}
\caption{The factor ${\mathcal N}$ playing the role of the reflection coefficient of the half space as a function  of momenta $\gamma$, $m=1, \lambda=1$.}
\label{F1}
\end{figure}

\begin{figure}[h]
\centering
\includegraphics[width=7cm]{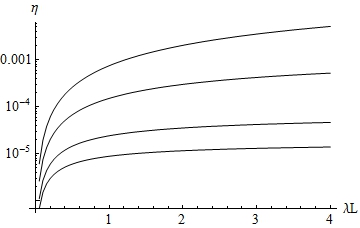}
\caption{The ratio, $\eta$, of the Casimir energy of Equation \Ref{a60} and the Casimir energy of the massless scalar field with Dirichlet boundary conditions on the plates. $\eta$  is drawn in logarithmic scale as a function of dimensionless separation $\lambda L$ for different vales of $\mu=m/\lambda$. From top to bottom, $\mu=0.001, 0.01, 0.5, 1$.}
\label{F2}
\end{figure}

In the foregoing sections, we considered the Casimir effect between two slabs in the framework of quantum field theory. A scalar field $\phi$ mimics the electromagnetic field, and another scalar field $\psi$, which is confined by Dirichlet boundary conditions, mimics the matter inside the slabs. Both fields interact by a Yukawa coupling. For the calculation of the vacuum interaction energy, we used the TGTG formula and calculated the reflection coefficient for the field $\phi$ from the one-loop polarization operator $\Pi$ of the field $\psi$. The polarization operator divides into a translational non-invariant part, $\Pi^{(nt)}$, and an invariant part, $\Pi^{(t)}$. While $\Pi^{(nt)}$ can be calculated in a straightforward manner, $\Pi^{(t)}$ has an ultraviolet divergence, which can be removed by standard methods of coupling renormalization. Together, the polarization operator, and with it the reflection coefficient, can be calculated numerically (see Figure \ref{F1}), and their asymptotics for large and small momenta can be obtained (Equations \Ref{a58} and \Ref{a59}). Finally, the Casimir energy can be calculated (see Figure \ref{F2}).
\vspace{-12pt}

As discussed in Section \ref{sec4}, the considered model has an instability that can be avoided by a~more realistic model with a momentum dependent coupling.
Thus, the main result of the paper is to demonstrate how the Casimir energy can be calculated for a (3 +1)-dimensional matter field in the slabs within the framework of quantum field theory beyond   cases with graphene where the matter field is (2 +1)-dimensional. Thus, the path to Casimir energy calculations for more realistic models of   matter is opened.

\vspace{6pt}



\begin{thebibliography}{99}
\bibitem[Casimir and Polder(1948)]{casi48-73-360}
Casimir, H.B.G.; Polder, D.
\newblock {The Influence of Retardation on the London-van der Waals Forces}.
\newblock {\em Phys.~Rev.} {\bf 1948}, {\em 73},~360, doi:10.1103/PhysRev.73.360.

\bibitem[Lifshitz(1956)]{lifs56-2-73}
Lifshitz, E.M.
\newblock The Theory of Molecular Attractive Forces Between Solids.
\newblock {\em J. Exp. Theor. Phys.} {\bf 1956}, {\em 2},~73--83.

\bibitem[Bordag \em{et~al.}(2009)Bordag, Klimchitskaya, Mohideen, and
  Mostepanenko]{BKMM}
Bordag, M.; Klimchitskaya, G.L.; Mohideen, U.; Mostepanenko, V.M.
\newblock {\em Advances in the Casimir Effect}; Oxford University Press: Oxford, UK, 2009.

\bibitem[Scheel and Buhmann(2008)]{sche08-58}
Scheel, S.; Buhmann, S.Y.
\newblock {Macroscopic Quantum Electrodynamics---Concepts and Applications}.
\newblock {\em {A}cta~{P}hys.~{S}lovaca} {\bf 2008}, {\em 58},~675, doi:10.2478/v10155-010-0092-x.

\bibitem[Philbin({2010})]{phil10-12-123008}
Philbin, T.G.
\newblock {Canonical quantization of macroscopic electromagnetism}.
\newblock {\em {New J. Phys.}} {\bf {2010}}, {\em {12}}, doi:10.1088/ 1367-2630/12/12/123008.

\bibitem[Horsley and Philbin({2014})]{hors14-16-013030}
Horsley, S.A.R.; Philbin, T.G.
\newblock {Canonical quantization of electromagnetism in spatially dispersive
  media}.
\newblock {\em {New J. Phys.}} {\bf {2014}}, {\em {16}}, 013030.

\bibitem[Bordag \em{et~al.}(2009)Bordag, Fialkovsky, Gitman, and
  Vassilevich]{bord09-80-245406}
Bordag, M.; Fialkovsky, I.V.; Gitman, D.M.; Vassilevich, D.V.
\newblock {Casimir interaction between a perfect conductor and graphene
  described by the Dirac model}.
\newblock {\em Phys.~Rev.~B} {\bf 2009}, {\em 80},~245406, doi:10.1103/PhysRevB.80.245406.

\bibitem[Pyatkovskiy(2009)]{pyat09-21-025506}
Pyatkovskiy, P.K.
\newblock {Dynamical polarization, screening, and plasmons in gapped graphene.}
\newblock {\em J.~Phys. Condens.~Matter} {\bf 2009}, {\em 21},~025506.

\bibitem[Fialkovsky and Vassilevich({2012})]{fial12-27-1260007}
Fialkovsky, I.V.; Vassilevich, D.V.
\newblock {Quantum field theory in graphene}.
\newblock {\em Int.~J.~Mod.~Phys.~A} {\bf {2012}}, {\em {27}},~{1260007}.

\bibitem[Zavialov(1990)]{Zavialov90}
Zavialov, O.
\newblock {\em Renormalized Quantum Field Theory}; Kluwer Academic Publishers:
  Dordrecht, The~Netherlands,~1990.

\bibitem[Berman \em{et~al.}({2014})Berman, Ford, and Milonni]{berm14-89-022127}
Berman, P.R.; Ford, G.W.; Milonni, P.W.
\newblock {Nonperturbative Calculation of the London-van der Waals Interaction
  Potential}.
\newblock {\em Phys.~Rev.~A} {\bf {2014}}, {\em {89}}, 022127.

\bibitem[Bordag(2017)]{bord17-96-062504}
Bordag, M.
\newblock {Casimir and Casimir--Polder forces with dissipation from first
  principles}.
\newblock {\em Phys.~Rev.~A} {\bf 2017}, {\em 96},~062504.

\bibitem[Martin and Rothen(2004)]{Martin2004}
Martin, P.; Rothen, F.
\newblock {\em Many-Body Problems and Quantum Field Theory: An Introduction}:
 Springer: Berlin/Heidelberg, Germany, 2004.
\end{thebibliography}
\end{document}